\documentclass[proof]{WileyASNA-v2}

\articletype{Article Type}%

\received{...}
\revised{...}
\accepted{...}

\begin{document}

\title{The long-term variability of a population of ULXs monitored by Chandra}

\author[1]{Hannah P. Earnshaw*}
\author[2]{Gauri Patti}
\author[1]{Murray Brightman}
\author[3]{Rajath Sathyaprakash}
\author[4]{Dominic J. Walton}
\author[5]{Felix Fürst}
\author[6]{Timothy P. Roberts}
\author[1]{Fiona A. Harrison}

\authormark{HANNAH EARNSHAW \textsc{et al}}

\address[1]{\orgname{California Institute of Technology}, Pasadena, CA, USA}
\address[2]{\orgname{Université Paris Sciences \& Lettres}, Paris, France}
\address[3]{\orgname{Istituto Universitario di Studi Superiori}, Pavia, Italy}
\address[4]{\orgname{Centre for Astrophysics Research, University of Hertfordshire}, Hatfield, UK}
\address[5]{\orgname{European Space Astronomy Centre (ESAC)}, Madrid, Spain}
\address[6]{\orgname{Centre for Extragalactic Astronomy, Durham University}, Durham, UK}

\corres{*Hannah Earnshaw. \email{hpearn@caltech.edu}}

\abstract{We present preliminary results of a {\it Chandra} Large Program to monitor the ultraluminous X-ray source (ULX) populations of three nearby, ULX-rich galaxies over the course of a year, finding the ULX population to show a variety of long-term variability behaviours. Of a sample of 36 ULXs, some show persistent or moderately variable flux, often with a significant relationship between hardness and luminosity, consistent with a supercritically accreting source with varying accretion rates. Six show very high-amplitude variability with no strong relationship between luminosity and hardness, though not all of them show evidence of any long-term periodicity, nor of the bimodal distribution indicative of the propeller effect. We find evidence of additional eclipses for two previously-identified eclipsing ULXs. Additionally, many sources that were previously identified as ULXs in previous studies were not detected at ULX luminosities during our monitoring campaign, indicating a large number of transient ULXs. 
}


\maketitle

\section{Introduction}\label{sec:intro}

Ultraluminous X-ray sources (ULXs) are the brightest X-ray sources found outside of galaxy centres, empirically defined as being non-nuclear X-ray sources with luminosities above 10$^{39}$\,erg\,s$^{-1}$, the Eddington luminosity of a $\sim$10\,M$_{\odot}$ black hole. ULXs are generally thought to be stellar-mass compact objects undergoing super-Eddington accretion (for a recent review, see \citealt{king23}). This has been confirmed for a handful of ULXs for which we have been able to detect X-ray pulsations \citep{bachetti14,fuerst16,israel17,carpano18,sathyaprakash19,rodriguezcastillo20}, demonstrating that the compact object is a neutron star (NS). Whether or not pulsations are detected, ULXs generally show similar spectra, consisting of two thermal components below 10\,keV and a steep power-law excess above 10\,keV (e.g. \citealt{walton18}), attributed to a super-Eddington accretion state involving a geometrically thick supercritical accretion disk, radiatively-driven outflowing winds, and the contribution from an accretion column in the case of NS accretors.

\begin{figure*}[t]
\centering
\includegraphics[height=5.6cm]{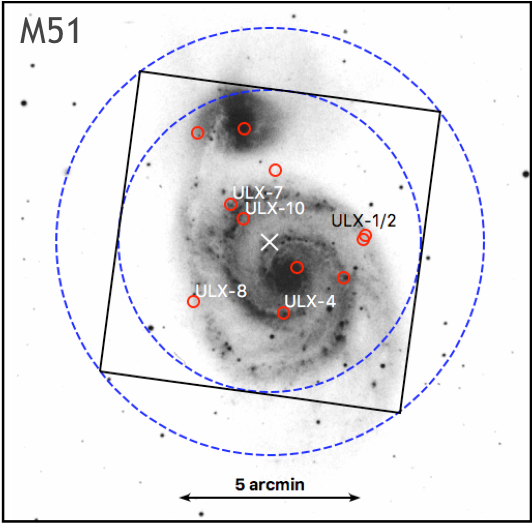}
\includegraphics[height=5.6cm]{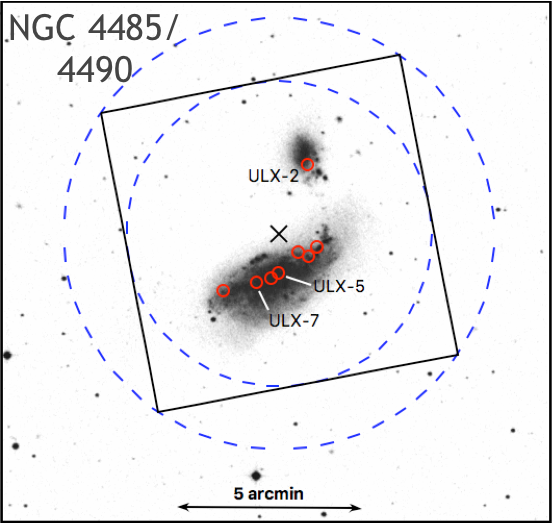}
\includegraphics[height=5.6cm]{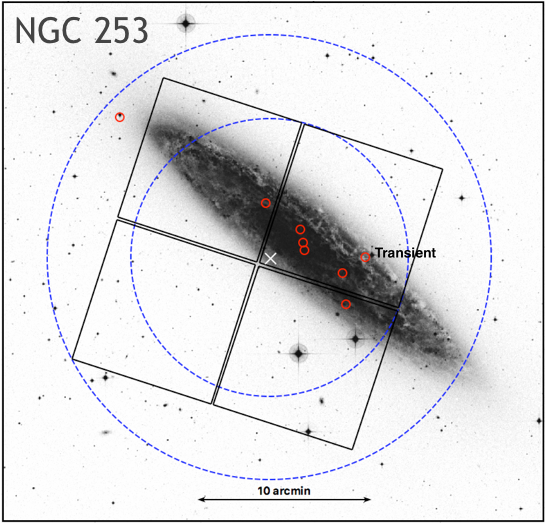}
\caption{Optical DSS images of our selected ULX host galaxies: M51, NGC 4485/4490, and NGC 253. The galaxies M51 and NGC 4485/4490 are shown with a single example ACIS-S detector chip orientation, and NGC 253 is shown with the same for ACIS-I. Aimpoints are marked with a cross, and the regions always within the {\it Chandra} field of view or possibly within the field of view (depending on telescope roll angle) are shown with the inner and outer blue dashed circles respectively. Known ULXs at the time of observation are indicated with red circles, and previously-studied sources of note are annotated. \label{fig:galaxies}}
\end{figure*}

Some ULXs show significant long-term variability in flux. For example, several bright, well-studied ULXs have demonstrated significant spectral variation over the time that they have been revisited (e.g. \citealt{walton17,walton20}), with variation in either or both thermal components potentially driving these changes. Several ULXs identified as NS accretors show apparent superorbital periodicity on the order of tens to hundreds of days (e.g. \citealt{brightman20}), due to precession of some part of the accreting system. NSs may also undergo dramatic drops in flux related to the propeller effect, and both effects may be present in the same source (e.g. \citealt{fuerst23}). Superorbital periodicities have also been found in a few ULXs for which pulsations have not been detected (e.g. \citealt{earnshaw22}) although without understanding the exact mechanism behind the periodicity, this is not by itself conclusive evidence of a NS accretor. Two ULXs have been found to show eclipses \citep{urquhart16}. All of these mechanisms can cause dramatic changes in flux between observations, including causing the source to leave and reenter the empirically defined ULX luminosity regime.

Our best understanding of long-term variability in ULXs so far comes from the long-term study of ULXs that are already interesting for being particularly bright or for demonstrating pulsations -- though these sources alone show a range of different behaviours (e.g. \citealt{gurpide21}). Many more ULXs are monitored on a regular basis using short snapshots taken with {\it Swift}-XRT, which can track variability within the ULX regime well for reasonably nearby sources. However, the ULX population as a whole is relatively sparsely observed to a depth sufficient to explore the large changes in flux that these sources can exhibit (e.g. \citealt{earnshaw18}).

This motivated performing multiple {\it Chandra} observations of three nearby galaxies to explore the variability of a sample of ULXs over the course of a year down to fluxes $\sim$10$^{38}$\,erg\,s$^{-1}$, allowing us to probe order-of-magnitude changes in flux even below the ULX luminosity regime. In these proceedings we share some preliminary results from this program. 

\section{Sample Selection \& Methods}\label{sec:sample}

In order to obtain a large sample of ULXs that could be observed with a small number of {\it Chandra} pointings, we identified three galaxies from the \citet{earnshaw19} catalogue that each contained multiple ULXs and whose ULX population could fit within the field of view of either one ACIS-S detector or the ACIS-I array. In this way, we selected the galaxy pairs M51 \citep{terashima04} and NGC 4485/4490 \citep{gladstone09}), and the nearby spiral galaxy NGC 253 \citep{wik14}. We observed each galaxy ten times each on an approximately monthly cadence depending on its visibility throughout the year, with all observations occurring between October 2020 and January 2022. We chose exposure times that would allow significant detection of sources at a luminosity of 10$^{38}$\,erg\,s$^{-1}$ (35\,ks per observation for M51 at 9.1 Mpc, 30\,ks for NGC 4485/4490 at 8.9 Mpc, and 10\,ks for NGC 253 at 3.4 Mpc). We show all three galaxies and their ULX populations known at the time of proposal, along with the {\it Chandra} aimpoints and example fields of view throughout the year, in Fig.~\ref{fig:galaxies}.

\begin{figure*}[t]
\centering
\includegraphics[width=8.78cm]{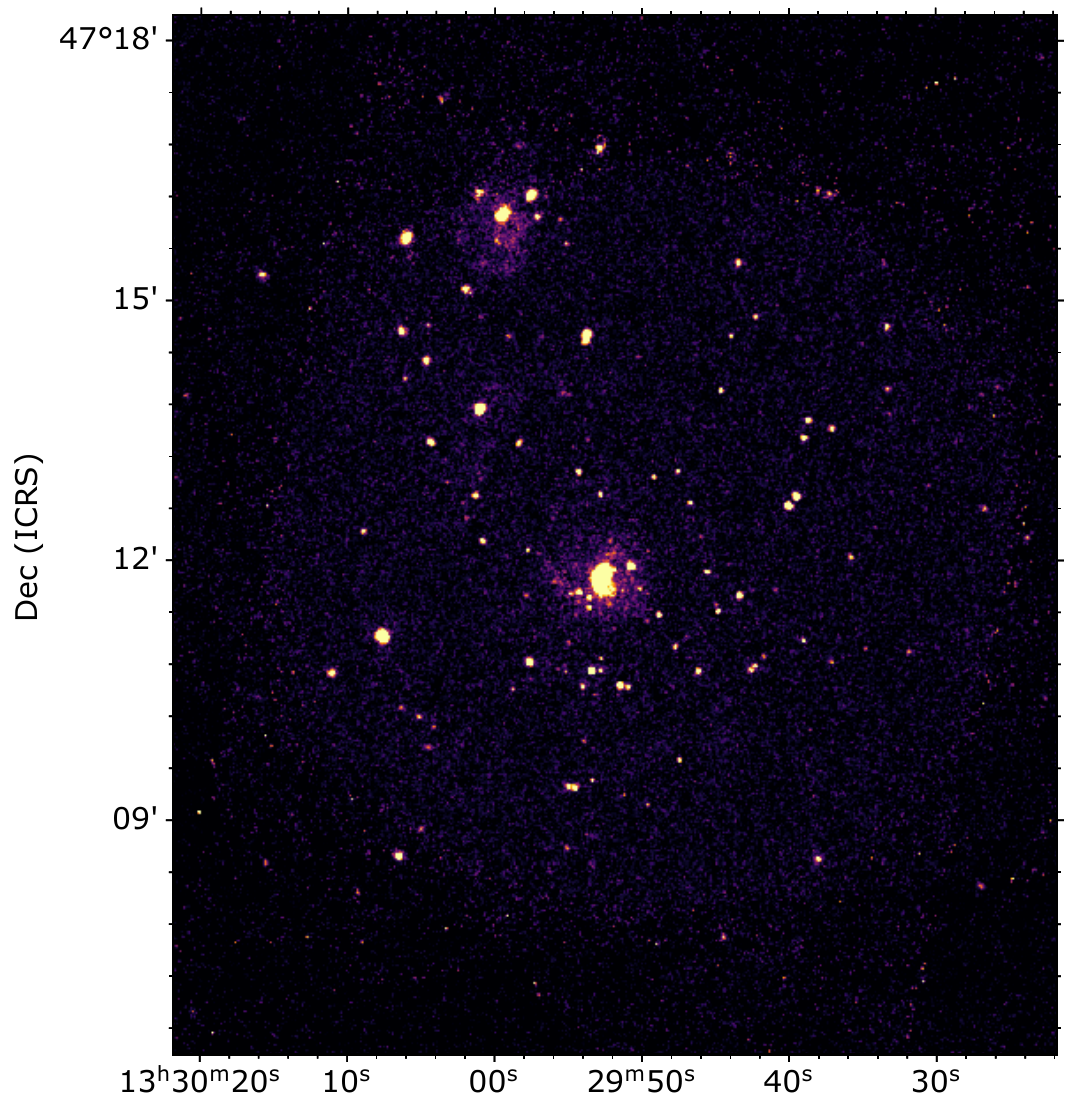} 
\includegraphics[width=8.4cm]{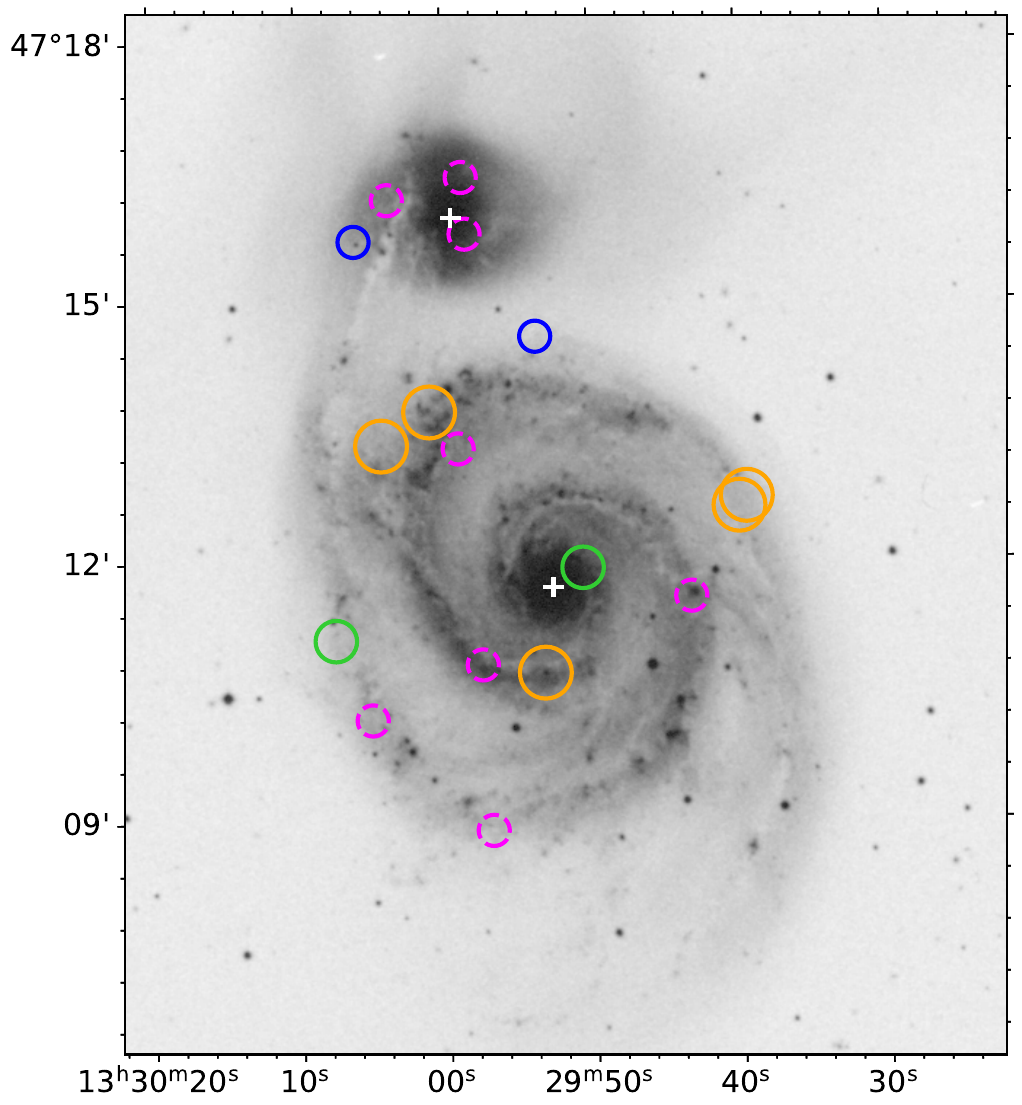} 
\includegraphics[width=8.78cm]{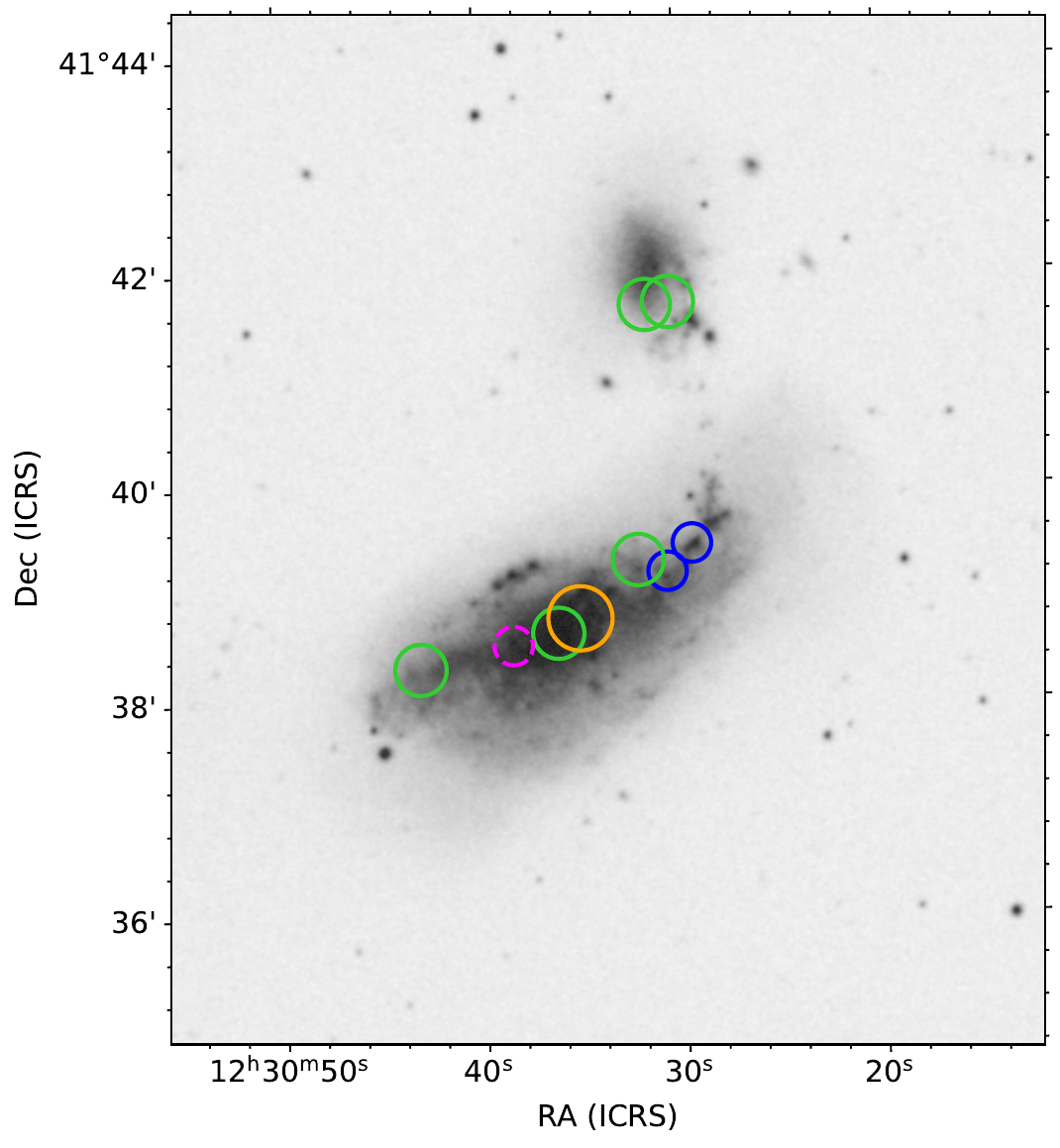}
\includegraphics[width=8.4cm]{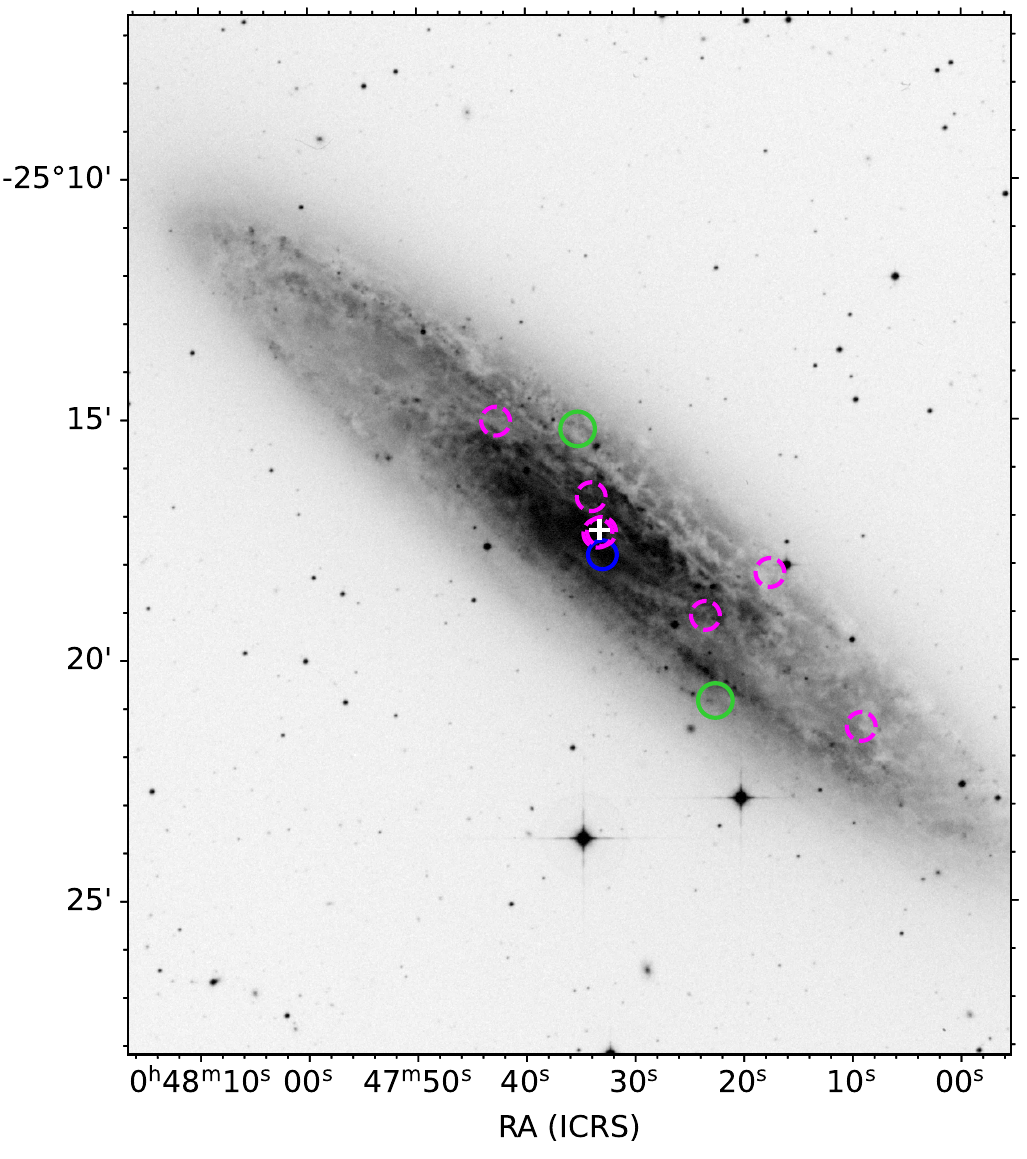}
\caption{\textit{Top left}, stacked {\it Chandra} image of M51 using all monitoring observations, showing a large number of bright X-ray point sources. \textit{Top right}, optical DSS image of M51 with the locations of ULXs from our sample marked in circles. Persistent ULXs are marked with small blue circles, moderately variable ULXs are marked with medium green circles, and highly variable ULXs are marked with large orange circles. Previously-identified ULXs that are not at ULX luminosities during our monitoring are marked with dashed magenta circles. AGNs are marked with white crosses. \textit{Bottom left}, optical DSS image of NGC 4485/4490. \textit{Bottom right,} optical DSS image of NGC 253. ULXs are marked using the same system as for M51. \label{fig:sources}}
\end{figure*}

We performed source detection on each observation using the CIAO routine {\tt wavdetect}. Sources were identified by matching 5$\sigma$ detections across all observations, then the routine {\tt srcflux} was used to calculate the flux or upper limit for every source in every observation, whether or not it was significantly detected. We defined our sample of ULXs first by selecting all sources with at least one detection with luminosity $>$10$^{39}$\,erg\,s$^{-1}$. We also added to our sample any further ULXs identified in previous investigations, for which we used the \citet{walton22} sample, which collates ULXs identified in the archival data of {\it XMM-Newton}, {\it Chandra}, and {\it Swift}. Altogether, this resulted in a sample of 36 sources that are known to have reached ULX luminosities at one point, two of which were observed at ULX luminosities for the first time during our monitoring campaign (M51~ULX-13, NGC~4485-90~ULX-9).

In order to study how the ULX spectra change with time, we used the Bayesian Estimation of Hardness Ratios (BEHR; \citealt{park06}) software, which is able to reliably produce hardness ratio estimates and uncertainties in the low-counts regime, including when a source is not detected in an energy band, making it particularly useful for our investigation in which we expect to explore a wide range of luminosities and therefore potentially low count rates. We calculated the hardness ratio as defined as $HR = (H-S)/(H+S)$, using soft band $S$ = 0.3--2\,keV, and hard band $H$ = 2--10\,keV. To validate this approach, we also used the {\tt specextract} routine to extract spectra for some of the bright sources in the sample. For these relatively short observations with {\it Chandra}, we did not have the data quality to fit a more complex model than an absorbed power-law, but we were able to confirm that variations in hardness ratio accurately traced variations in power-law photon index and thus the general spectral shape. 

\begin{figure*}[t]
\centering
\includegraphics[width=\textwidth]{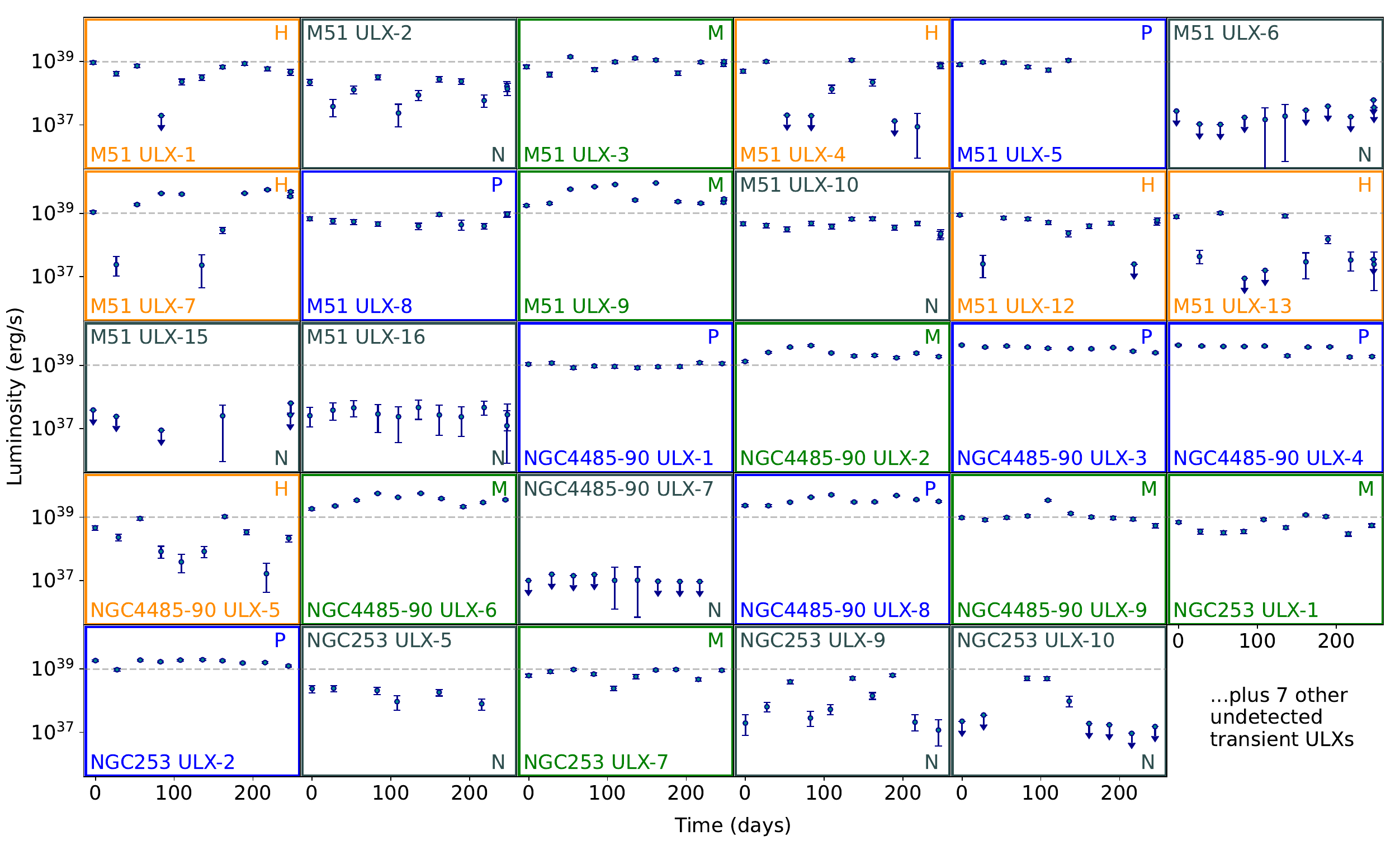}
\caption{The light curves for each ULX in our sample with at least one {\it Chandra} detection during our monitoring, labelled with the source name and whether we classified the source as persistent (blue, labelled P), moderately-variable (green, M), highly-variable (orange, H), or not a ULX during our monitoring (grey, N). Seven further previously-observed ULXs were not detected at all during our monitoring campaign.  \label{fig:lightcurves}}
\end{figure*}

\section{Results \& Discussion}\label{sec:results}

We divided our sample of ULXs into four subsamples based on the degree of variability they exhibit. We defined Persistent ULXs as those whose luminosity varied by less than a factor of three over the course of our monitoring {\it Chandra} observations, Moderately Variable ULXs as those which varied by a factor of between three and ten over the course of our observations, and Highly Variable ULXs as those that varied by over a factor of ten. We also defined a subsample of sources that have been identified as ULXs at other times, but do not achieve a luminosity greater than 10$^{39}$erg\,s$^{-1}$ in any of our {\it Chandra} observations. Of the 36 ULXs in our sample, we found seven to be persistent, seven moderately variable, six highly variable, and 16 that did not reach ULX luminosities during our monitoring campaign. We show the locations of all ULXs and their categorisations in Fig.~\ref{fig:sources}.

For each ULX that was detected at least once during our monitoring campaign, we plotted a light curve of its luminosity over the time that the host galaxy was monitored (Fig.~\ref{fig:lightcurves}). For sources that we classified as persistent or moderately-variable, we see fairly similar behaviour only with differing intensities -- most sources appear to show gradual, irregular changes in luminosity over time, with some showing a shift of a factor of a few from one observation to another (though these changes could still be happening on a timescale of $\sim$weeks given the monitoring cadence). Interestingly, if we include all archival {\it Chandra} data, most of these sources retain their variability classification with only three showing a higher range of variability over the longer timescale, though we note that previous {\it Chandra} observations of these galaxies do not provide the regular sampling we were able to achieve with our campaign. 

The highly-variable ULXs show a variety of behaviours, though all but one of them only reach ULX luminosities at their highest fluxes. Some appear to demonstrate potential periodicity in their variability, which may indicate the presence of a superorbital period, although caution is warranted in this case since our limited sampling is only able to place an upper limit on a potential long-term periodicity. For example, while the light curve for M51 ULX-7 appears to demonstrate a $\sim$100-day periodicity, we already know this source to have a 44-day superorbital period \citep{brightman22} which our monitoring cadence is too coarse to fully sample. 

\begin{figure*}[t]
\centering
\includegraphics[width=\textwidth]{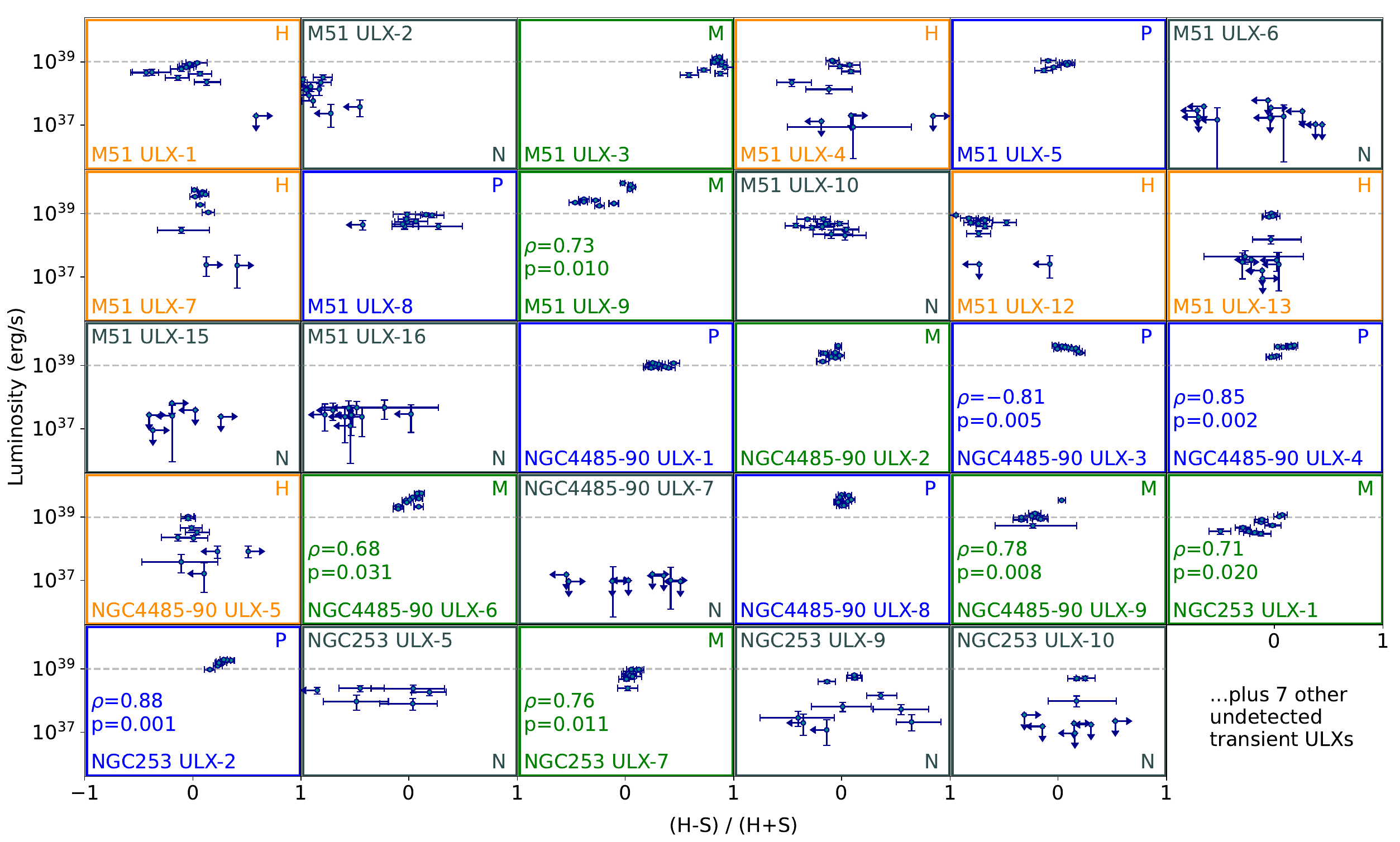}
\caption{The hardness ratio against luminosity for each ULX in our sample with at least one {\it Chandra} detection. Labels are as described in Fig.~\ref{fig:lightcurves}. For sources with a significant correlation between hardness and luminosity, we also label the Pearson's coefficient $\rho$ and the associated $p$-value. \label{fig:hr}}
\end{figure*}

We also note that the large drops in M51 ULX-1 and M51 ULX-12 (both identified as eclipsing ULXs under the names ULX-2 and ULX-1 respectively in \citealt{urquhart16}) may be consistent with additional eclipses for these sources. While we do not see any evidence for an ingress or egress in the individual observation light curves, the $\sim$35\,ks observations are shorter than the known lower limit of the eclipse durations (minimum of $\sim$90\,ks for ULX-12 and $\sim$48\,ks for ULX-1). None of the other highly-variable ULXs show behaviour consistent with eclipses (i.e. a single observation at a time with a dramatically lower flux than the average).

Among the sources previously identified as ULXs that did not reach ULX luminosities during our monitoring, we also see several different behaviours. Seven were not detected at all, and three others (M51 ULX-6, M51 ULX-15, and NGC4485/4490 ULX-7) only have a couple of low-significance flux measurements, which may be contaminated by extended emission or low-luminosity source confusion and so may not be reliable detections. Of these `disappeared' ULXs, many are likely to have been transient events, such as XT-1 in M51 \citep{brightman20}, which underwent a bright outburst and then faded away, and has not been detected since. 

Some other sources that are no longer ULXs at the time of our monitoring are still bright and consistently detected but with luminosities below 10$^{39}$\,erg\,s$^{-1}$. In particular, NGC 253 ULX-9 and ULX-10 exhibit qualitatively similar behaviour to the highly-variable sample of ULXs, except that their maximum luminosities are below the 10$^{39}$\,erg\,s$^{-1}$ threshold. Given how close to this threshold most of the other highly-variable ULXs are at their maximum luminosities, it is plausible that these two sources have variability mechanisms similar to the rest of the highly-variable ULX subsample.

To further explore the spectral change with luminosity for these sources, we plot the the hardness ratio against luminosity for our ULX sample in Fig.~\ref{fig:hr}, which shows a wide variety of different behaviours. While a handful of ULXs are very hard or very soft, most ULXs of all variability types sit more-or-less in the middle of the hardness range, indicating significant contributions from both the hot and cool thermal components. 

We performed an initial search for correlations between the hardness and luminosity by calculating the Pearson's coefficient for each source. Significant correlations and their $p$-value are annotated on Fig.~\ref{fig:hr}. We find that eight ULXs across our sample show a significant correlation: three of the persistent sample and five of the moderately-variable sample, with all but one showing a positive correlation with the hardness increasing with the luminosity. None of the highly-variable sample show a significant correlation between hardness and luminosity. This may be affected by the higher degree of uncertainty in hardness at the lower fluxes we observe, but even on visual inspection these sources are either mostly consistent with there being no significant change in hardness with luminosity, or display more complex behaviour than a simple, well-defined trend.

A positive relation between luminosity and hardness may be expected from a super-Eddington accreting system, in which an increased mass accretion rate would be expected to increase the spherization radius of the wind \citep{poutanen07}, which would result in a cooler soft component. In the case of a system viewed at low inclinations down the funnel of the wind, an increased mass accretion rate may also increase the disk scale height and/or the characteristic beaming radius, leading to an increase in geometrically beamed emission from the hot, supercritical disk (e.g. \citealt{walton17}; though this may also lead to a softening of the hot component). In the case of a negative correlation between luminosity and hardness, a source viewed at higher inclinations may grow softer with high mass accretion rate if the opening angle of the funnel decreases and more of the outflowing wind component crosses the line of sight and obscures more of the hot inner component (dramatic increases in obscuration would lead to an overall decrease rather than an increase in flux, though the negative correlation we see does not span a large range of luminosities). 

These results appear to be fairly consistent with the findings of \citet{gurpide21}, who also find more generally positive hardness-luminosity correlations. They also find that the spectral variability of ULXs tends to be more complex than a straightforward trend between hardness and luminosity, showing, for example, a degeneracy at lower luminosities between a straightforward decrease in mass accretion rate and an abrupt increase in obscuration by an optically thick wind at high mass accretion rates. It is possible that we see evidence of such a degeneracy in our own sample in e.g. M51 ULX-3, as well as in some of sources for which we still find a significant correlation.

We know the variability of M51 ULX-7 is due to a high-amplitude superorbital periodicity \citep{brightman20,brightman22}, and the lack of strong spectral variation over the course of its periodicity rules out periodic obscuration by a warped accretion disk. Its high levels of variability likely originate from the precession of some part of the system -- the inner accretion disk, the outflowing wind, or the magnetic dipole itself, since we know the source to be a neutron star. However, similar explanations can't necessarily be applied to other highly-variable ULXs such as M51 ULX-13 that also have no particular change in hardness despite large changes in luminosity, since there is also no evidence of superorbital periodicity. Furthermore, given the wide range of luminosities observed for the highly-variable ULXs that are not known to be eclipsing, dramatic drops in flux due to the propeller effect do not seem to be a viable cause either. Further investigation is required to determine the cause of the high levels of variability we observe in these cases. 

\section{Conclusions \& Future Work}\label{sec:conc}

As a population, ULXs show a wide variety of variability behaviour, from persistent luminosity and slow changes to dramatic changes in flux on timescales of weeks, which may be attributed to multiple physical mechanisms. The observed ULX population is very changeable, with the sources exhibiting ULX luminosities in any single observation of a galaxy only a fraction of the total population of sources that reach those luminosities over the course of a year, even when not considering the larger number of transient ULXs. Infrequent visits to X-ray-source-rich galaxies may not be able to capture the true population of extremely-accreting sources within. 

A full paper on this investigation is in preparation containing further analysis of this population, including more detailed timing analysis such as the production of covariance spectra, full spectral studies of the bright sources in the population, and more in-depth comparisons with the hardness ratio properties of other known ULXs. 

\section*{Acknowledgments}

HPE acknowledges acknowledges support under NASA grant GO1-22049X and NASA contract NNG08FD60C. The majority of this work was performed on the traditional homeland of the Tongva people.

\bibliography{cmproceedings}

\end{document}